\documentclass[aps,prc,twocolumn,floatfix]
{revtex4-1}
\usepackage{amsfonts}
\usepackage{amsmath}
\usepackage{amssymb}
\usepackage{graphicx}
\usepackage{rotating}
\usepackage{verbatim}
\usepackage{hyperref}
\usepackage{color}   
\usepackage[dvipsnames]{xcolor}
\usepackage{soul}

\begin{document}

\title{Phase structure of magnetized 2SC quark matter under compact star conditions}

\author{M. Coppola$^{1,2}$}
\author{P. Allen$^{1}$}
\author{A.G. Grunfeld$^{1,2}$}
\author{N.N. Scoccola$^{1,2,3}$}

\affiliation{$^{1}$Department of Theoretical Physics, Comisi\'on Nacional de Energ\'ia
At\'omica, Av.Libertador 8250, 1429 Buenos Aires, Argentina~\\
 $^{2}$CONICET, Rivadavia 1917, 1033 Buenos Aires, Argentina~\\
 $^{3}$Universidad Favaloro, Sol\'is 453, 1078 Buenos Aires, Argentina}


\begin{abstract}
Some properties of magnetized two flavor color superconducting (2SC) cold dense quark matter under compact star conditions (COSC) are investigated within a $SU(2)_f$ Nambu Jona-Lasinio type model. We study the phase diagram for several model parametrizations. The features of each phase are analyzed through the behavior of the chiral and superconducting condensates for increasing chemical potential or magnetic field. We show how the phases are modified in the presence of $\beta$-equilibrium as well as color and electric charge neutrality conditions.
\end{abstract}

\maketitle
\section{Introduction}
During the last years many works in the literature were devoted to the study of strongly interacting matter under the influence of strong magnetic fields (see e.g.~\cite{Kharzeev:2012ph} and refs. therein).
In the astrophysics scenario, this is motivated by the fact that certain compact objects called magnetars can have surface magnetic fields up to $10^{15}$ G~\cite{Duncan:1992hi}, with estimates for the magnetic field values at their centers of a few orders of magnitude larger~\cite{Chatterjee:2014qsa}. There, the relevant region of the QCD phase diagram is that of low temperature and intermediate values of density, where color superconducting (CSC) phases of quark matter are expected to exist. In this regime lattice QCD (LQCD) calculations are hindered by the well-known sign problem.
Effective models of QCD arise then as a powerful tool to circumvent these problems. One of these models is the Nambu Jona-Lasinio (NJL) model~\cite{reports} in which quarks are assumed to interact locally.
The local character of the interactions leads to divergences in the momentum integrals that need to be regularized. When the magnetic field is introduced, the vacuum energy acquires a Landau level (LL) structure and additional care is required in the treatment of the divergences. A convenient regularization method is the so-called ``Magnetic Field Independent Regularization'' (MFIR) scheme~\cite{Menezes:2008qt,Allen:2015paa}, where unphysical oscillations (that occur in other regularization schemes) are completely removed. In this contribution we present a brief report on the analysis of the behavior of magnetized 2SC quark matter under color and electric charge neutrality as well as $\beta$-equilibrium, which will be referred to as COSC, in the framework of NJL-type model using the MFIR scheme. Further details of this work can be found in Ref.~\cite{Coppola:2017edn}.

\section{The model and its regularization} \label{SecII}

We consider a Lagrangian density composed of a quark sector, described by a NJL-type $SU(2)_{f}$ Lagrangian density which includes scalar-pseudoscalar and color pairing interactions, and a leptonic contribution.  In the presence of an external magnetic field and finite chemical potentials it reads
\begin{eqnarray}
\mathcal{L} & = & \bar{\psi}\left[i\ \tilde{\rlap/\!D}-m_{c}+\hat{\mu} \gamma^{0}\right]\psi\nonumber +G\left[\left(\bar{\psi}\psi\right)^{2}+\left(\bar{\psi}i\gamma_{5}\vec{\tau}\psi\right)^{2}\right] \\
&& +H\left[(i\bar{\psi}^{C} \epsilon_{f}\epsilon_{c}^{3}\gamma_{5}\psi)(i\bar{\psi} \epsilon_{f}\epsilon_{c}^{3}\gamma_{5}\psi^{C})\right] \nonumber\\
&& +\sum_{l=e,\mu}  \bar{\psi}_{l}\left[i\gamma^{\mu}\left(\partial_{\mu}-ieA_{\mu}\right)-m_{l}+\text{\ensuremath{\mu}}_{l}\gamma^{0}\right]\psi_{l}.
\label{lagr}
\end{eqnarray}
Here, $G$ and $H$ are coupling constants, $\psi\!=\!\left(u,d\right)^{T}$ represents a quark field with two flavors, $\psi^{C}\!=\!C\bar{\psi}^{T}$ and $\bar{\psi}^{C}\!=\!\psi^{T}C$, with $C\!=\!i\gamma^{2}\gamma^{0}$, are charge-conjugate spinors and $\vec{\tau}=(\tau_{1},\tau_{2},\tau_{3})$ are Pauli matrices. Moreover, $(\epsilon_{c}^{3})^{ab}\!=\!(\epsilon_{c})^{3ab}$ and $(\epsilon_{f})^{ij}$ are antisymmetric matrices in color and flavor space respectively. Note that we adopt the usual 2SC ansatz $ \Delta_5= \Delta_7=0$, $\Delta_2= \Delta$, where the blue quark is unpaired~\cite{reports}. The (current) quark mass $m_{c}$ is  taken as equal for both up and down flavors, while $m_e\!=\!0.511\:\mathrm{MeV}$ and $m_\mu\!=\!105.66\:\mathrm{MeV}$.
The presence of an external magnetic field is introduced through the covariant derivative $\tilde{D}_{\mu}=\partial_{\mu}-i\tilde{e}\tilde{Q}\tilde{{\cal A}}_{\mu}$, where $\tilde{A}_\mu$ is a massless rotated $U(1)$ field~\cite{Alford:1999pb}. In units of $\tilde{e}$ the associated rotated charges are: $\tilde{q}_{ub}\!=\!1$, $\tilde{q}_{ur}\!=\!\tilde{q}_{ug}\!=\!1/2$, $\tilde{q}_{dr}\!=\!\tilde{q}_{dg}\!=\!-1/2$ and $\tilde{q}_{db}\!=\!0$. In the present work we consider a static and constant magnetic field in the 3-direction, $\tilde{{\cal A}}_{\mu}=\delta_{\mu2}x_{1}B$. In order to describe the system as a function of $\tilde{e}B$, we will take $\tilde{e}=e\cos\theta \simeq e$ and $\tilde{A}_\mu \simeq A_\mu$ as a good approximation based on the small estimated value of $\theta$~\cite{Gorbar:2000ms}.

In what follows we work in the mean field approximation (MFA), assuming that the only non-vanishing expectation values are $<\bar{\psi}\psi>=-(M-m_{c})/2G$ and $<i\bar{\psi}^{C} \epsilon_{f}\epsilon_{c}^{3}\gamma_{5}\psi>=-\Delta/2H$, which can be chosen to be real. Here, M and $\Delta$ are the so-called dressed quark mass and superconducting gap, respectively.

The quark chemical potential is introduced through the diagonal matrix $ \hat{\mu}=(\mu_{ur},\mu_{ug},\mu_{ub},\mu_{dr},\mu_{dg},\text{\ensuremath{\mu}}_{db}) =
\mu+Q\mu_{Q}+T^{8}\mu_{8}$. Here $\mu$ is the common chemical potential for non-zero baryonic density while $\mu_{8}$ and $\mu_{Q}$ are added to ensure color and electric charge neutrality conditions respectively. The six quantities $\mu_{fc}$ are in principle independent parameters, but become related among themselves under COSC. On one hand, the quarks paired by the interaction are degenerate. Since for the chosen ansatz red and green quarks are paired, their densities will be equal and we can impose $\mu_{3}=0$, which implies $\mu_{ur}=\mu_{ug}$ and $\mu_{dr}=\mu_{dg}$. On the other, assuming that no neutrinos are trapped in the system, $\beta$-equilibrium conditions lead to $\mu_{\mu}\!=\!\mu_{e}$ and $\mu_{dc}\!=\!\mu_{uc}+\mu_{e}$, where the latter implies $\mu_{Q}\!=\!-\mu_{e}$. For calculational simplicity, it is also convenient to define
\begin{equation}
\bar{\mu} \!=\! \frac{\mu_{dg}+\mu_{ur}}{2}, \quad
\delta\mu \!=\! \frac{\mu_{dg}-\mu_{ur}}{2}\!=\!\frac{1}{2}\mu_{e}.
\label{relmu2}
\end{equation}

The resulting MFA thermodynamic potential at vanishing temperature reads
\begin{equation}
\Omega_{\mbox{\scriptsize{MFA} }}=\frac{(M-m_{c})^{2}}{4G}+\frac{\Delta^{2}}{4H}-\sum_{\left|\tilde{q}\right|=0,\frac{1}{2},1}\!P_{\left|\tilde{q}\right|}-P_{lep}.\label{omegaMFA}
\end{equation}
The pressures appearing in Eq.~(\ref{omegaMFA}) read
\begin{eqnarray}
P_0 & = & \int\frac{d^{3}p}{(2\pi)^{3}}\big(E_{db}^{+}+\left|E_{db}^{-}\right|\big),\label{omega0} \\
P_1 & = & \frac{\tilde{e}B}{8\pi^{2}}\sum_{k=0}^{\infty}\alpha_{k}\int_{-\infty}^{\infty}dp_{z}\big(E_{ub}^{+}+\left|E_{ub}^{-}\right|\big),\label{omega1}\\
P_{1/2} & = & \frac{\tilde{e}B}{8\pi^{2}}\sum_{k=0}^{\infty}\alpha_{k}\int_{-\infty}^{\infty}dp_{z} \sum_{\lambda,s=\pm}\!E_{\Delta^s}^\lambda,\label{omega12} \\
P_{lep}&=&\sum_{l=e,\mu} P_1\Big\rvert_{M=m_l, \:\:\mu_{ub}=\mu_l} \,. \label{omegalep}
\end{eqnarray}
Here, we have introduced $\alpha_{k}=2-\delta_{k0}$ and
\begin{eqnarray}
E_{db}^{\pm} & = & \sqrt{p^{2}+M^{2}}\pm\mu_{db},\\
E_{ub}^\pm & = & \sqrt{p_{z}^{2}+2k\tilde{e}B+M^{2}}\pm\mu_{ub}, \label{E1}\\
E^\lambda_{\Delta^s} & = & \sqrt{\left(\sqrt{p_{z}^{2}+k\tilde{e}B+M^{2}}+\lambda\bar{\mu}\right)^{2}+\Delta^{2}}+s\delta\mu\,. \quad \label{E12}
\end{eqnarray}

To regularize the divergent expressions in Eqs.~(\ref{omega0}-\ref{omegalep}) we use the MFIR scheme, where the divergence is removed by subtracting a vacuum term with the form of the $\tilde{e}B\!=\!0$ case. The remaining magnetic field-dependent contributions turn out to be finite, and the vacuum terms are regularized with a sharp 3D cutoff~\cite{Menezes:2008qt,Allen:2015paa}. It is worth mentioning that when $P_{1/2}$ is regularized, a term proportional to $\Theta(\delta\mu-\Delta)$ appears. One can therefore distinguish between two possible situations within a $\Delta \neq 0$ phase: a ``gapless phase'' ($g2SC$) when $\delta\mu>\Delta$, and an ordinary $2SC$ phase when $\Delta>\delta\mu$. The source of the differences between these phases comes from the changes in the quasi-particle spectrum in Eq.~(\ref{E12}). As explained in Ref.~\cite{Huang:2003xd}, when $\delta\mu\neq0$ the gap equation has two branches of solutions and the modes are no longer completely degenerate, but they split into pairs of two with gaps $\Delta_{\pm}=\Delta\pm\delta\mu$. While in the $2SC$ phase the four modes are gapped, in the $g2SC$ phase the lower dispersion relation for the quasi-particle crosses the zero-energy axis and two of the four modes become gapless.

Given the regularized form of Eq.~(\ref{omegaMFA}), $\Omega_{\mbox{\scriptsize{MFA}}}^{reg}$, the minimum for fixed values of $\mu$ and $\tilde{e}B$ is found by solving the gap equations subject to the neutrality conditions
\begin{equation}
\frac{\partial\Omega_{\mbox{\scriptsize{MFA}}}^{reg}}{\partial\xi}=0,\qquad\xi=M,\Delta,\mu_8,\mu_e.\label{gapeq}
\end{equation}
The density of each particle species is obtained by deriving the thermodynamical potential with respect to the corresponding chemical potential.

We will consider two $SU(2)_{f}$ NJL model parametrizations. They are listed in Table~\ref{Table1}, where $M_{0}$ represents the vacuum effective quark mass in the absence of external magnetic fields.

\begin{table}[h]
\caption{\label{Table1} Parameter sets for the SU(2)$_{f}$ NJL model. In both cases, empirical values in vacuum for the pion observables are reproduced,
$m_{\pi}=138\:\mathrm{MeV}$ and $f_{\pi}=92.4\:\mathrm{MeV}$.}
\vspace{0.2cm}
\centering{}%
\begin{tabular}{c|c|c|c|c|c}
\hline
\hline\hspace*{0.2cm} \hspace*{0.2cm}  & \hspace*{0.2cm} $M_{0}$ \hspace*{0.2cm} & \hspace*{0.2cm} $m_{c}$ \hspace*{0.2cm}  & \hspace*{0.2cm} $G\Lambda^{2}$ \hspace*{0.2cm}  & \hspace*{0.2cm} $\Lambda$ \hspace*{0.2cm}  & \hspace*{0.2cm} $-<u\bar{u}>^{1/3}$ \hspace*{0.2cm} \tabularnewline
 & MeV  & MeV  & - & MeV  & MeV \tabularnewline
\hline
Set~1  & 340  & 5.59  & 2.21  & 621  & 244 \tabularnewline
Set~2  & 400  & 5.83  & 2.44  & 588  & 241 \tabularnewline
\hline
\end{tabular}
\end{table}

\section{Numerical Results} \label{SecIII}

We consider the ratio $H/G\!=\!0.75$, which is favored by various models of the quark effective interaction. We find the presence of $B$, $D$, and $A$ phases, where  $D$ and
$A$ can in turn find themselves in a g2SC or a 2SC mode. Their
main characteristics are summarized in Table~\ref{Table2}. Note
that MC and IMC stand for ``Magnetic
Catalysis''\cite{Shovkovy:2012zn} and ``Inverse Magnetic
Catalysis''\cite{Preis:2012fh}, respectively. Meanwhile, vA-dH
refers to ``van Alphen-de Haas'' transitions.

\begin{table}[h]
\caption{\label{Table2} Characteristics and convention of each phase.}
\vspace{0.2cm}
\centering{}%
\begin{tabular}{c|c}
Phase & Characteristics
\tabularnewline
\hline
$B$ - Vacuum &
\begin{tabular}{@{}c@{}}
$\chi$-symmetry broken, $M\!=\!M(B,\mu\!=\!0)$, \\
MC, low $\mu$, $\mu\!<\!M$, $\Delta\!=\!\mu_8\!=\!\mu_e\!=\!n\!=\!0$
\end{tabular}
 \tabularnewline
 \hline
$A$ - CSC &
\begin{tabular}{@{}c@{}}
$\chi$-symmetry almost restored, IMC, \\
big $\mu$, ($\Delta,\mu_8,\mu_e,n)\!\neq\!0$, vA-dH transitions
\end{tabular}
\tabularnewline \hline
$D$ - Mixed &
\begin{tabular}{@{}c@{}}
$M_B>M_D>M_A$, IMC, \\
$(\Delta,\mu_e,n)_A\!>\!(\Delta,\mu_e,n)_D\!>\!
(\Delta,\mu_e,n)_B$
\end{tabular}
\tabularnewline
- - - - - - - - &
- - - - - - - - - - - - - - - -
- - - - - - - - - - - - -
\tabularnewline
\begin{tabular}{@{}c@{}}
2SC \\ g2SC-Gapless
\end{tabular}
& \begin{tabular}{@{}c@{}}
$\Delta_\pm>\delta\mu$, four gapped, $n_{dr}=n_{ur}$ \\ $\Delta_\pm<\delta\mu$, two gapless, $n_{dr}\neq n_{ur}$
\end{tabular}
\tabularnewline
& \upbracefill
\tabularnewline
& \begin{tabular}{@{}c@{}}
Present in D and A. Doubly-degenerate \\  modes with gaps $\Delta_\pm=\Delta \pm \delta\mu$
\end{tabular}
\end{tabular}
\end{table}

\begin{figure}
\centering{}\includegraphics[width=\columnwidth]{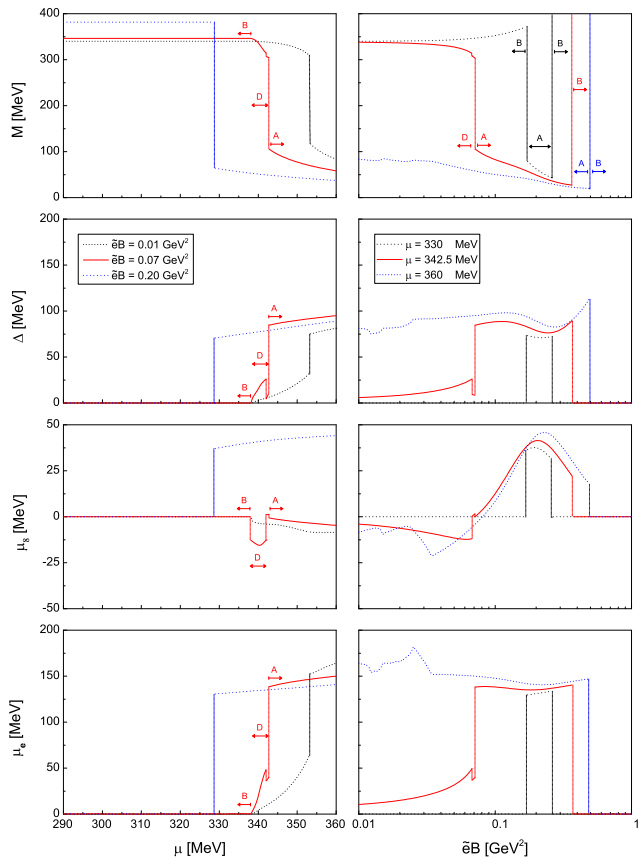}
\caption{(Color online) $M$, $\Delta$, $\mu_{8}$ and $\mu_{e}$  for $H/G=0.75$ within Set~1 considering three representative fixed values. Left: vs $\mu$ for fixed $\tilde{e}B$. The $A,B,D$ labels correspond to $\tilde{e}B=0.07\:\mathrm{GeV}^2$.  Right: vs $\tilde{e}B$ for fixed $\mu$.} \label{Fig1}
\end{figure}

In the left panel of Fig.~\ref{Fig1} we show the results for $M$, $\Delta$, $\mu_e$
and $\mu_8$ as a function of $\mu$ for three representative values
of $\tilde{e}B$ within Set~1. For clarity, the different phases are labeled for the $\tilde{e}B=0.07\:\mathrm{GeV}^2$ case. Near the transition to the
superconducting $A$ phase, $M$ takes values around
$100\:\mathrm{MeV}$ and diminishes toward a value slightly above
$m_c$ for higher chemical potentials. As for $\Delta$ and
$\mu_{e}$, they increase with $\mu$ in the range considered and
can both acquire high values. In general, $\Delta$ can reach values of the order of 100 MeV, while $\mu_e<200\:\mathrm{MeV}$ and $\mu_8$ will lie in the
range $\left|\mu_{8}\right|<30\:\mathrm{MeV}$, varying between
positive and negative values. The transition from $B$ or $D$ to
$A$ is of first order, while the transition from $B$ to $D$ is of
second order. We should also note that $\mu_8$ appears to be
discontinuous along this transition.
However, we bear in mind that in the $B$ phase its value is
actually not well-defined, and since it is arbitrarily taken to be
zero, such discontinuity has no physical meaning.

Superconductivity is suppressed with respect to the case without COSC~\cite{Allen:2015paa}. The presence of $\mu_e$ separates the Fermi momenta of the up and down quarks with respect
to each other. Since the quark pairing occurs between particles of equal and opposite momenta, this splitting reduces the diquark condensate. It can be interpreted that, due to this suppression, the $D$ phase exists because the $A$ phase with larger $\Delta$ is energetically disfavored with respect to the former. 

For $\tilde{e}B=0.07\; \mathrm{GeV^{2}}$, we see that immediately after
the second order transition from $B$ to $D$, $\Delta\gtrsim\delta\mu$, so the system is in the $2SC$ mode. However, these two quantities are very similar, and when the chemical potential is increased $\delta \mu$ becomes larger than $\Delta$ for $\mu\simeq342\;\mathrm{MeV}$, leading to a $g2SC$ region.

We turn now to the right panel of Fig.~\ref{Fig1}.
We find that, in the $B$ phase, $M$ always increases with the magnetic field according to the MC effect, which occurs principally in vacuum. Particulary, this can be seen in the figure for the $\mu=330$ MeV case.
In the $D$ phase, $M$ decreases slowly
with $\tilde{e}B$, exhibiting the behavior of IMC. In the $A$ phase we see a series of near vertical first order
transitions. They are related to the vA-dH effect and their origin 
lies in the quantization in LL's of the dispersion relations of 
quarks and leptons under the influence of magnetic fields, according 
to Eq.~(\ref{E1}). An important point is that no
oscillations of the parameters are present in the $B$ phase, in
contrast to previous studies which use smooth regularization
functions~\cite{Fayazbakhsh:2010gc,Mandal:2012fq}. This is because
the MFIR scheme removes these strong unphysical oscillations, also
assuring that all of the oscillations present in the $A$ phase are
real vA-dH transitions.

\begin{figure}[h]
\centering{}\includegraphics[width=\columnwidth]{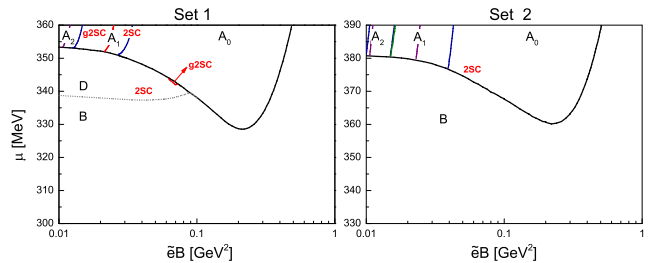}
\caption{(Color online) Phase diagrams for $H/G=0.75$, for both Set~1 (left) and Set~2 (right). Full lines correspond to first order transitions and dotted lines to second order transitions. Black lines represent phase transitions. VA-dH transitions for unit charged species are colored as: blue for $ub$ quarks, violet for electrons and green for muons. Red lines indicate $g2SC$-$2SC$ transitions.}
\label{Fig2}
\end{figure}

The corresponding phase diagrams in the $\tilde{e}B-\mu$ plane are
given in Fig.~\ref{Fig2} for the two parameter sets considered. Since Set~2 corresponds to a higher value of $M_0$ the transitions are
displaced to higher chemical potentials. For Set~2  only phases $B$ and $A$ exist. In Set~1, instead, phases $B$, $D$ and $A$ are present, where the latter two can exist both in the $g2SC$ or the $2SC$ modes. We will concentrate on Set~1 in what follows. The behavior as a
function of $\tilde eB$ of the first order transition leading to the $A$-type phases is worth noting: for small fields, the
transition has a small downward slope, which becomes sharper as
the magnetic field is increased, forming a well-shaped curve. This
effect is related to the so-called IMC effect~\cite{Preis:2012fh}.
From comparison with~\cite{Allen:2015paa}, COSC tend to decrease the IMC effect:
the depth of the well decreases.

The $B$ phase is present for low chemical potentials ($\mu \lesssim 330\: \mathrm{MeV}$). Moreover, in the MFIR scheme it is
always recovered for high enough magnetic fields. Regarding the $D$ phase, its presence depends on the set of parameters. The existence of this intermediate mixed phase is a consequence of the diquark pairing alone and hence is already present for zero magnetic field, extending itself in the horizontal direction.

In the $A$ phase we see the presence of the already mentioned vA-dH transitions. They are similar to those discussed in~\cite{Allen:2015paa}. However, in the 2SC case with a rotated magnetic field under COSC only the $ub$ quarks and leptons exhibit
ordinary vA-dH transitions as in the basic NJL model. The $db$
quarks are not coupled to the field, so the form of the dispersion
relation is that of the free quark. As for the paired $r$ and $g$
quarks, all of their LL's are usually populated when $\Delta \neq 0$.
Only in the g2SC phase the $P_{1/2}$ term may contain a sum
over LL's which is cut off by a Heaviside function. However, the
origin of this transitions is different from that of the vA-dH
ones. The effect of these transitions is smeared out as $H/G$ is 
increased. Since they are numerous and become rather weak to affect 
the order parameters in a visible way, we have not included them in 
the phase diagram. In the $A_i$ sub-phases of Fig.~\ref{Fig2}, the $ub$ quark populates up to the $i$-th LL. If we traverse the phase diagram horizontally in the
increasing $\tilde eB$ direction, the highest populated LL of a
particular unit charged species decreases in one unit every time
one of the corresponding transitions is crossed. In the $A_0$
phase, both $ub$ quarks and leptons are in the LLL. The maximum LL 
populated in these phase diagrams is smaller than the one obtained in: 
(a) the symmetric matter case, because $\mu_e$ reduces $\mu_{ub}$, and (b) the
model without superconductivity, because $\tilde{q}_{ub} > q_u,|q_d|$. 
Note that for $H/G=0.75$, in the $D$ phase there is a small gapless region,
bounded by first order transitions. From analyzing the $B=0$ case,
this region is expected to grow rapidly when $H/G$ is decreased
between $0.75$ and $0.7$.

Although the ratio $H/G\!=\!0.75$ is favored by various models of the quark effective interaction, from a more phenomenological point of view this value is subject to rather large uncertainties. For this reason, another representative value of this ratio, $H/G\!=\!1$, was considered in~\cite{Coppola:2017edn}. In that case, superconducting effects are enhanced and $\Delta$ can reach values of the order of 200 MeV. This will induce an increase in the density of paired quarks. To maintain color neutrality, $\mu_8$ is lowered, lying in the range $\left|\mu_{8}\right|<70\:\mathrm{MeV}$, while $\mu_e$ is only slightly modified.

\section{Conclusions}
\label{SecIV}

As expected, several effects take place on the
behavior of the cold magnetized 2SC quark matter when introducing COSC. Their presence induces the existence of
$g2SC$ and $2SC$ modes, and reduces the maximum Landau level
reached in the vA-dH transitions. In general, charge neutrality
constraints tend to reduce the superconducting effect, increasing
the value of the critical chemical potentials and attenuating the
magnetic catalysis effect, diminishing the depth of the IMC well. As a consequence, the phase diagram is moved to higher ranges of $\mu$ favoring different phases. 
In this work we analyzed the usual $H/G=0.75$ ratio. For higher ratios superconducting effects are enhanced: $\Delta$ is increased, the  phase diagram is brought to lower ranges of $\mu$ and the IMC well widens and shortens~\cite{Coppola:2017edn}. For values of $H/G \gtrsim 1$ (which are unlikely to be realized in QCD) the
structure of the phase diagram is expected to be maintained, with the difference that the first order transition
between the D and A phases weakens and eventually turns into a crossover around $H/G \sim 1.15$. In particular, for $H/G<0.65$ it is expected that the superconducting gap vanishes in the $C$ and $A$ phases.

\section*{Acknowledgements}

This work has been funded in part by CONICET (Argentina) under
Grant No. PIP 578 and by ANPCyT (Argentina) under Grant No.
PICT-2014-0492.


\end{document}